\documentclass[aps, amsmath, 
11pt,
notitlepage,
superscriptaddress
]{revtex4-1}
\usepackage{graphicx, hyperref, cleveref, placeins, xcolor}
\graphicspath{{./figures/}}

\begin{document}

\title{Plasmonic gain in current biased tilted Dirac nodes}
\author{Sang Hyun Park}
\affiliation{Department of Electrical \& Computer Engineering, University of Minnesota, Minneapolis, Minnesota, 55455, USA}
\author{Michael Sammon}
\affiliation{Department of Electrical \& Computer Engineering, University of Minnesota, Minneapolis, Minnesota, 55455, USA}
\author{Eugene Mele}
\affiliation{Department of Physics and Astronomy, University of Pennsylvania, Philadelphia, Pennsylvania, 19104, USA}
\author{Tony Low}
\email{tlow@umn.edu}
\affiliation{Department of Electrical \& Computer Engineering, University of Minnesota, Minneapolis, Minnesota, 55455, USA}

\begin{abstract}
	Surface plasmons, which allow extreme confinement of light, suffer from high intrinsic electronic losses. It has been shown that stimulated emission of electrons can transfer energy to plasmons and compensate for the high intrinsic losses. To-date, these realizations have relied on introducing an external gain media coupled to the surface plasmon. Here, we propose that plasmons in two-dimensional materials with closely located electron and hole Fermi pockets can experience gain, when an electrical current bias is applied along the displaced electron-hole pockets, without the need for an external gain media. As a prototypical example, we consider WTe$_2$ from the family of 1T$'$-MX$_2$ materials, whose electronic structure can be described within a type-II tilted massive Dirac model. We find that the nonlocal plasmonic response experiences prominent gain for experimentally accessible currents on the order of mA$\mu$m$^{-1}$. Furthermore, the group velocity of the plasmon found from the isofrequency curves imply that the amplified plasmons are highly collimated along a direction perpendicular to the Dirac node tilt when the electrical current is applied along it. 
	\end{abstract}

\maketitle

\section{Introduction}

Surface plasmons\cite{Maier2007} are collective excitations of electrons that result in highly confined electromagnetic modes in metals or semiconductors. 
In two-dimensional materials\cite{Low2017, Basov2016}, such as graphene\cite{Jablan2009, Low2014b}, propagating surface plasmons have been observed\cite{Fei2012, Chen2012, Dai2015} and are accompanied by a variety of interesting phenomena such as plasmonic waveguiding\cite{Vakil2011, Liu2011a}, topological plasmons\cite{Jin2017, Jung2018, Jin2016b}, chiral directional plasmons\cite{Song2016, Kumar2016a}, and the plasmonic Fizeau drag effect\cite{Dong2021a, Zhao2021, Borgnia2015,VanDuppen2016}, among many others. High quality factor surface plasmons in any two-dimensional material is key to the observation of these near-field plasmonic phenomena. However, the intrinsically high electronic losses of the plasmons is a significant barrier to their performance and applicability\cite{Ni2018, Boltasseva2011}. Hence, it is highly desirable to overcome the fundamental limitation on quality factor set by the electronic losses. One route to overcoming this limitation is to introduce enough gain to compensate for the intrinsic losses\cite{Xiao2010a}.

Gain, or wave amplification, is a phenomenon that is ubiquitous in nature.
Rogue waves, Rijke tubes, and optical parametric amplification are all examples in which a wave is amplified rather than damped as it propagates through a medium\cite{Rijke1859,RAYLEIGH1878,Dysthe2008,Solli2007, Baumgartner1979}. In most of these examples, the amplification is driven by external pumping, such as a heater or laser that can inject energy into the system. Early studies to impart gain to plasmons began with the realization that stimulated emission can also amplify plasmons in a way similar to amplification of light\cite{Bergman2003a}. The first experimental demonstrations were able to amplify plasmons by coupling to external media that were either optically\cite{Oulton2009, Noginov2009} or electrically\cite{Hill2009} pumped, realizing plasmonic nanolasers with extremely compact field confinement. Optically pumped setups\cite{Nezhad2010, Lu2012} require gain medium to be adjacent to the surface plasmon mode. Electrically pumped systems encapsulate a typical semiconductor laser system in conjunction with a plasmonic mode in a metal-insulator-metal gap, accompanied by population inversion within the insulating layer. 
However, integrating these approaches into on-chip nanophotonics has been challenging due to difficulties in down-sizing the optical gain components and issues related to contact resistance of small metallic structures\cite{Azzam2020, Ding2013}.

 Here, we propose that plasmons in a two-dimensional material with separated electron and hole pockets can experience internal plasmonic gain in the presence of an applied electrical current. The plasmonic medium derives its gain from nonlocal single particle transitions between the electron and hole bands and does not require an external gain medium. The electrons(holes) acquire a momentum anti-parallel(parallel) to the applied current direction. The shifted electron and hole distributions open up a finite energy momentum phase space window where population inversion can be achieved for single-particle transitions. When this phase space overlaps with the plasmon dispersion, plasmonic gain can occur. Such a condition for gain can be satisfied in two-dimensional materials with low carrier density when the electron and hole pockets have a separation on the order of the plasmon momentum. Examples of materials that meet these prerequisites are found in the 1T$'$-MX$_2$ family of transitional metal dichalcogenides such as WTe$_2$ and MoTe$_2$\cite{Qian2014a, Soluyanov2015}.

\section{Results}
The inverted band structure found in many 1T$'$-MX$_2$\cite{Qian2014a, Soluyanov2015} materials follows the tilted 2D massive Dirac model\cite{Sodemann2015, Du2018}
\begin{equation}\label{eq:dirac_model}
	H=t\eta k_x + v(k_y\sigma_x + k_x \sigma_y) + (m/2-\alpha k^2)\sigma_z
\end{equation}
where $t$ is the tilt along the $k_x$ direction, $\eta=\pm 1$ is the node index, $v$ is the Fermi velocity, $m$ is the gap size, and $\alpha$ is a term that closes the Fermi surface at large $k$. This minimal model captures most of the salient features of the unique inverted band structure. The energy dispersion of \cref{eq:dirac_model} given by $\varepsilon_\pm(\mathbf{k})=t\eta k_x \pm [v^2 k^2 + (m/2 -\alpha k^2)^2]^{1/2}$ shows an interesting dependence on the tilt angle parameterized by the ratio $t/v$ (see \cref{fig:1}a). For $t/v<1$ the model describes a type-I Dirac node which has a single electron or hole pocket at Fermi energies in the conduction or valence band. For $t/v>1$ we find a type-II Dirac node which can support both an electron and hole Fermi surface where $\hat{\mathbf{k}}_{e-h}$, the relative displacement vector of the electron and hole pockets, is along the tilt direction. Here, we will be focusing on the type-II Dirac node with $E_F=0$eV which has electron and hole pockets of equal size. For definiteness, the parameters of the Dirac model are set to fit the tight-binding band structure of monolayer WTe$_2$\cite{Hu2021a} and are given as $v=2.427$eV\AA, $t=1.16v$, $m=0.138$eV, $\alpha=7.541$eV\AA$^2$.

Gain in the plasmons can be seen from a nonlocal description of the electron response. Hence we calculate the loss function  $L(\mathbf{q},\omega)=\textrm{Im}(1/\epsilon(\mathbf{q},\omega))$ which reveals the plasmon excitations of the system. The dielectric response is calculated from the random phase approximation to be $\epsilon(\mathbf{q},\omega)=1-v_c\Pi(\mathbf{q},\omega)$ where $v_c$ is the Coulomb potential and $\Pi(\mathbf{q},\omega)$ is the non-interacting polarizability. The polarizability is given by a sum of the contributions from two oppositely tilted nodes ($\Pi=\Pi^+ + \Pi^-$) where the polarizability of each node is given by

	\begin{equation}
		\Pi^\eta(\mathbf{q},\omega) = \frac{g_s}{4\pi^2}\int d^2\mathbf{k}\sum_{s s'}\frac{f(E_{s\mathbf{k}}^\eta)-f(E^{\eta}_{s'\mathbf{k}'})}{E^{\eta}_{s\mathbf{k}}-E^{\eta}_{s'\mathbf{k}'}+\hbar\omega+i\delta}|\langle\psi^{\eta}_{s\mathbf{k}}|\psi^{\eta}_{s'\mathbf{k}'}\rangle|^2.
	\end{equation}	
Here, $f(E_\mathbf{k})$ is the Fermi distribution, $\mathbf{k}'=\mathbf{k}+\mathbf{q}$, $E^{\eta}_{s\mathbf{k}}$ and $|\psi^{\eta}_{s\mathbf{k}}\rangle$ are the eigenvalues and eigenvectors of the tilted 2D massive Dirac Hamiltonian, and $g_s$ is the spin degeneracy. 
A DC current bias $\mathbf{j}$ applied to the system may then be modeled by modifying the Fermi distribution to $f_u(E_\mathbf{k}) = \left(e^{(E_\mathbf{k}-\mathbf{u}\cdot\mathbf{k}-E_F)/k_B T}+1\right)^{-1}$\cite{Bistritzer2009} where $\mathbf{u}$ is a parameter induced by the current density. The modified carrier occupation in valley $\eta=+1$ for a current applied along the positive $x$-direction is shown in \cref{fig:1}a. The applied electrical current requires that the electron and hole pockets in the $\eta=+1$($\eta=-1$) node are shifted closer to(away from) each other (see inset of \cref{fig:1}a). 

The loss function for a drift current $u/v=0.41$ and $\mathbf{q}=q_x\hat{\mathbf{x}}$ is shown in \cref{fig:1}b. We first note that the plasmon dispersion is skewed in the direction opposite to the drift current and exhibits a very prominent non-reciprocal behavior. Moreover, the loss function in the $-q_x$ plasmon branch becomes negative signifying gain in the plasmon. 
The origin of gain can be understood by examining the imaginary part of the polarizability from the $\eta=+1$ node shown in \cref{fig:1}c. The bias creates two regions in which the imaginary part of the polarizability is negative (blue regions labeled intra and inter in \cref{fig:1}c). Both these regions originate from single particle transitions from a higher energy to lower energy, i.e. an inverted population. The intraband transitions  occur mainly at large wavevectors of order $k_F$ and small frequencies that are distant from the plasmonic spectrum. In contrast, the interband transitions between the electron and hole pockets occur at much smaller wave vectors and much higher frequencies. As the bias is increased, the interband gain region can overlap with the plasmonic spectrum and impart gain. Indeed this can be seen by examining the boundary for interband transitions (see dashed magenta line in \cref{fig:1}b and c) connecting states above and below $E_F=0$ which matches well with the onset of plasmonic gain as shown in \cref{fig:1}b. Therefore we conclude that the observed plasmonic gain can be attributed to nonlocal interband transitions.

Note that the plasmonic gain as shown by the loss function in \cref{fig:1}a does not perfectly match the gain phase space suggested by the polarizability in \cref{fig:1}c. Specifically, the positive $q$ branch of the plasmon is not experiencing any gain despite overlapping with the interband gain phase space. This can be explained by taking into account the transitions from both the $\eta=1$ and $\eta=-1$ valleys. \cref{fig:1}b has 4 regions that are separated by the interband transition boundaries of each valley. Region 1 has no interband transitions while regions 2 and 3 have transitions from $\eta=+1$ and $\eta=-1$ respectively. The plasmon branch that resides in region 2 therefore experiences the gain from single particle interband transitions in the $\eta=+1$ valley. Plasmons in region 4 are influenced by interband transitions from both valleys, and since the $\eta=-1$ valley transitions only contribute to loss (see SI), gain from the $\eta=+1$ valley is cancelled out. 

\begin{figure}
	\centering
	\includegraphics[width=\textwidth]{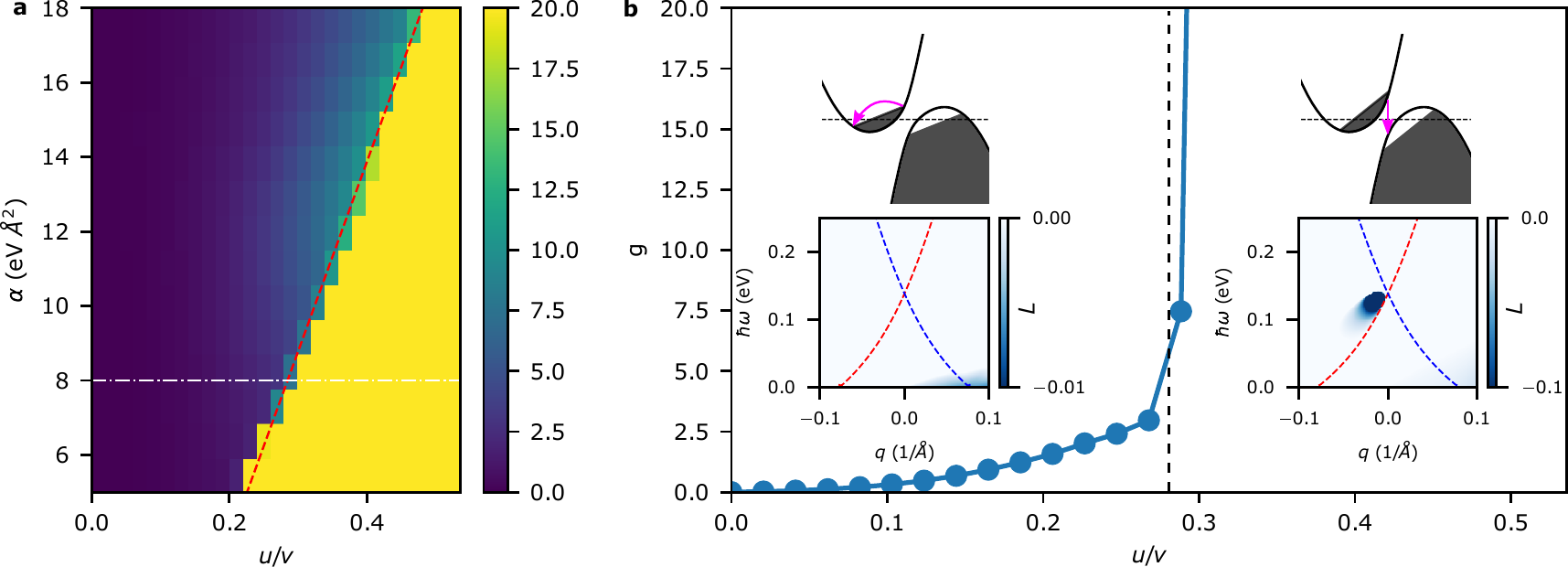}
	\caption{%
	\textbf{Threshold drift velocity for gain.}
	\textbf{a,} The gain metric $g(\alpha, u)$ is shown as a 2D color plot. A clear linear boundary between regions with and without plasmonic gain is observed. The red dashed line is a linear fitting to the transition boundary that gives $\alpha=50.99u/v-6.508$.
	\textbf{b,} A slice of the 2D color plot at $\alpha=8$eV\AA$^2$ (white dashed line in \cref{fig:2}a) is shown. The inset band structures schematically show the transitions responsible for $u/v=0.165$ and $u/v=0.33$ which are below and above $u_{th}$ respectively. The inset loss functions are scaled to only represent the regions over which $g(\alpha,u)$ is integrated as defined by $g(\alpha, u)=\left|\int_{L<0}L(q,\omega)dqd\omega\right|$. A black dashed line is drawn at $u_{th}/v=0.285$.}
	\label{fig:2}
\end{figure}

Having shown that plasmonic gain can indeed emerge for the tilted Dirac system, we now determine the threshold $u_{th}$ required for gain as a function of the model parameters. Since $\alpha$ closes the Fermi surface and therefore controls the size of the electron and hole pockets, we explore $u_{th}$ as a function of $\alpha$ while leaving all other parameters fixed to the values used for \cref{fig:1}. A larger $\alpha$ will in general lead to a smaller Fermi pocket size. Let us first define the metric for quantifying the gain in the system as
\begin{equation}\label{eq:gain_metric}
	g(\alpha, u)=\left|\int_{L<0}L(q,\omega)dqd\omega\right|
\end{equation}
where the integral is performed only over the region with gain ($L<0$). Results for $g(\alpha,u)$ are shown in \cref{fig:2}a. A clear linear boundary is observed between regions with and without plasmonic gain (a detailed analysis of the linear phase boundary is given in the supplementary information). Since the shift in the Fermi surface is proportional to both the parameter $u$ and the unbiased $k_F$, we can expect that a smaller Fermi surface will require a larger $u$ to reach the gain threshold. The threshold current density, which is dependent on both $u$ and the carrier density, is found to be on the order of a few mA$\mu$m$^{-1}$ and also decrease as a function of $\alpha$ (see SI). To take a closer look at what happens at the onset of gain, $g(\alpha,u)$ as a function of $u$ at $\alpha=8$eV\AA$^2$ is shown in \cref{fig:2}b. For $u<u_{th}$, the slow increase in gain is due to intraband transitions as we can see from the loss function (inset). For $u>u_{th}$, there is a sharp increase in gain which originates from the onset of interband transitions imparting gain to the plasmon. Note the difference in scale between the two loss function insets which explains why such an abrupt transition is observed for $g(\alpha,u)$.
We now expand our description of the plasmon to a general $\mathbf{q}=(q_x,q_y)$. At a given frequency, the complex plasmon wavevector $\mathbf{q}_{pl}=\mathbf{q}_{pl}'+i\mathbf{q}_{pl}''$ can be found by solving for the complex roots of $\epsilon(\mathbf{q},\omega)=0$. For small $\mathbf{q}''$, the complex roots are found by solving
\begin{subequations}
	\begin{gather}
	\textrm{Re}\left[\epsilon(\mathbf{q}'_{pl})\right]=0 \label{eq:pl_dispersion_re}\\
	 \mathbf{q}''_{pl}\cdot\left[\nabla_\mathbf{q'}\textrm{Re}(\epsilon)\right]_{\mathbf{q}'_{pl}}=-\textrm{Im}\left[\epsilon(\mathbf{q}_{pl}')\right].\label{eq:pl_dispersion_im}
	\end{gather}
\end{subequations}
 Note, however, that \cref{eq:pl_dispersion_im} does not uniquely define the direction of $\mathbf{q}_{pl}''$. $\mathbf{q}_{pl}''$ is taken to be parallel to the group velocity as the gain or loss is expected to be in the direction of the energy flow. The isofrequency curves for $u/v=0$ and $u/v=0.41$ are shown in \cref{fig:3}. As observed for the loss function in \cref{fig:1}, a finite drift velocity breaks reciprocity and induces gain in the plasmon spectrum. The isofrequency curves shown in \cref{fig:3}c,d further reveal that both the gain experienced by the plasmon for non-zero drift velocity and the topology of the isofrequency contour evolve with frequency. We find that at low frequency the isofrequency contour is separated into two distinct pockets, and that the amplified plasmons have a group velocity primarily along $\hat{\mathbf{k}}_{e-h}$. As the frequency is increased above $\hbar\omega=0.15$eV, these pockets merge into a single closed isofrequency surface which is relatively `flat' in the region of gain and has a group velocity perpendicular to $\hat{\mathbf{k}}_{e-h}$. Hence, highly directional propagation of the amplified plasmons perpendicular to $\hat{\mathbf{k}}_{e-h}$ is expected.

\begin{figure}
	\centering
	\includegraphics[width=\textwidth]{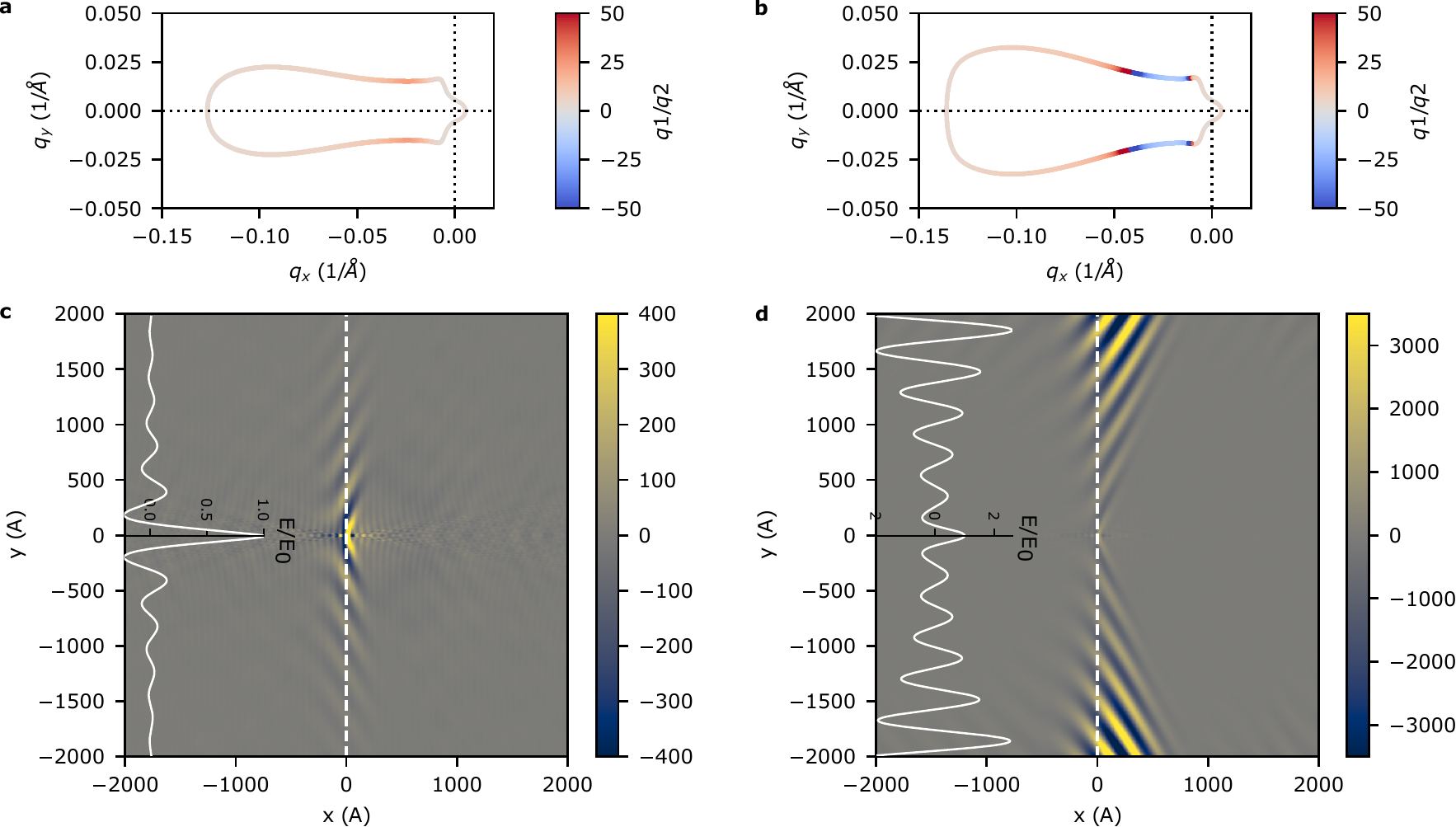}
	\caption{%
	\textbf{Fields excited by a point dipole.}
	\textbf{a,b,} Isofrequency curves at $\hbar\omega=0.16$eV for drift velocity $u/v=0.37$ and $u/v=0.41$ respectively. 
	\textbf{c,d,} Fields generated by \cref{eq:pw_superposition} for isofrequency curves \cref{fig:4}a and b respectively. The solid white inset plots show the field amplitude  along the white dashed lines and is normalized to the field at the origin.}
	\label{fig:4}
\end{figure}

With the complex plasmon wavevectors found, we now examine the real space behavior of the plasmons launched by a point source. The plasmon field can be written as a superposition of unidirectional plane-waves
\begin{equation}\label{eq:pw_superposition}
	\mathbf{E(r)}\propto \sum_{\mathbf{q}_{pl}}e^{i\mathbf{q}'_{pl}\cdot\mathbf{r}}e^{-\mathbf{q}''_{pl}\cdot\mathbf{r}}\Theta(\mathbf{v}_g\cdot\mathbf{r}).
\end{equation}
$\Theta$ is a step function that describes a one-sided plane wave propagating in the direction of the group velocity. Results for the plane wave superposition for $\hbar\omega=0.16$eV with drift velocities $u<u_{th}$ and $u>u_{th}$ are shown in \cref{fig:4}. For both cases, the plasmon propagation direction is roughly along the $y$-axis as expected from the isofrequency plots. It is important to emphasize that this is different from what one would expect from the phase velocity, which lies primarily in the $-q_x$-direction. For $u>u_{th}$, amplification of the plasmon in the direction of $\mathbf{v}_g$ is clearly observed. 

\section{Discussion \& Conclusion}
\begin{figure}
	\centering
	\includegraphics[width=\textwidth]{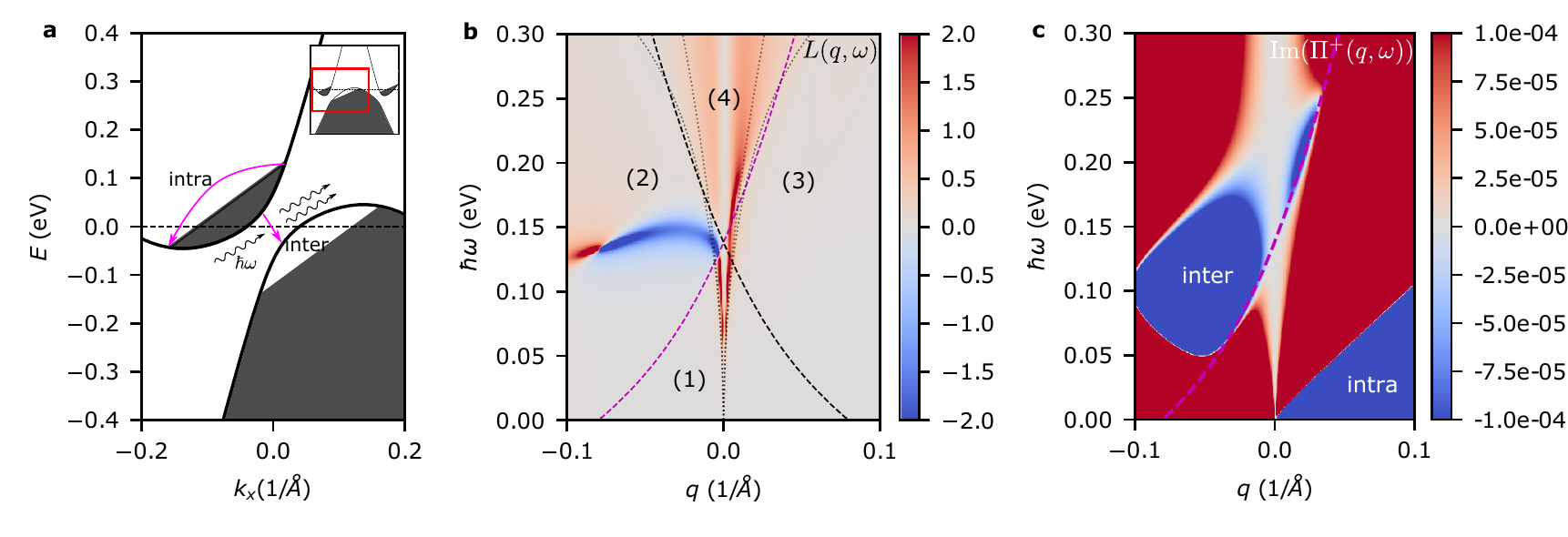}
	\caption{%
	\textbf{Tilted 2D massive Dirac model and loss function calculation.} 
	\textbf{a,} A representative energy band structure of \cref{eq:dirac_model} for the $\eta=+1$ node with a schematic representation of intraband and interband single particle transitions. Parameters fitted to the WTe$_2$ tight-binding band structure are $v=2.427$eV\AA, $t=1.16v$, $m=0.138$eV, $\alpha=7.541$eV\AA$^2$. We have further set $E_F=0$eV and $u=1$eV\AA. Inset shows the tight-binding band structure of the XTe$_2$ family where the red boxed region is the region we are modeling by the Dirac model.
	\textbf{b,} Loss function $L(q_x,\omega)$ calculation for parameters given above. The magenta(black) dashed lines show the interband transition boundaries for the $\eta=+1$($\eta=-1$) valley.
	\textbf{c,} Polarizability of the $\eta=+1$ valley. The two regions with negative $\textrm{Im}\Pi^+$ originate from intra and interband transitions.}
	\label{fig:1}
\end{figure}

\begin{figure}
	\centering
	\includegraphics[width=\textwidth]{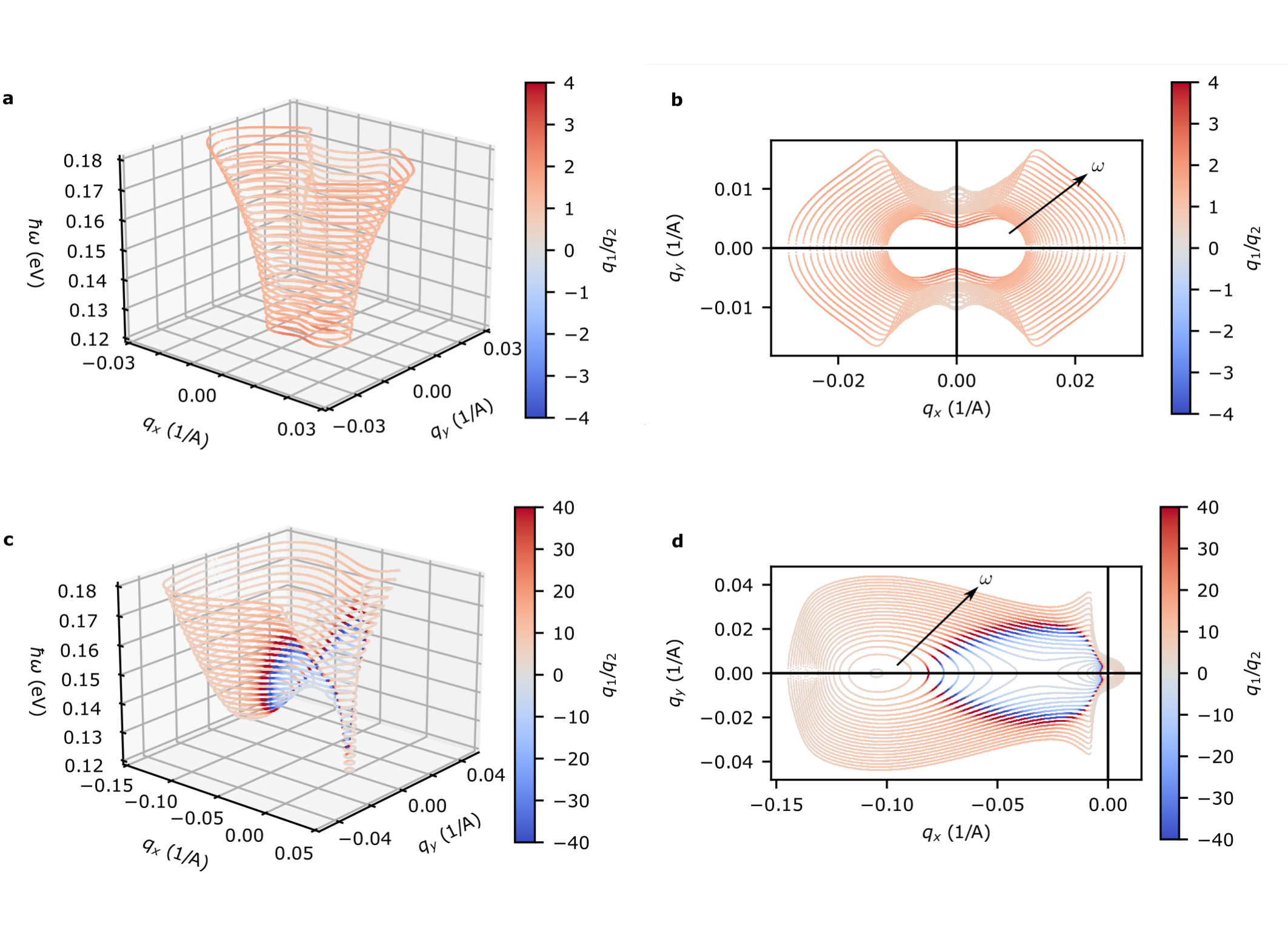}
	\caption{%
	\textbf{Isofrequency curves for the plasmon dispersion.}
	\textbf{a,} Isofrequency curves for $u=0$. All other parameters for the Dirac model are identical to parameters used in \cref{fig:1}. Color scale quantifies the quality factor $q_1/q_2$ which gives the number of cycles the plasmon goes through before decaying or amplifying by $e$. 
	\textbf{c,} Isofrequency curves for $u=1$eV\AA ($u/v=0.41$). Negative values of the quality factor denote regions of the plasmon dispersion that experience gain.
	\textbf{b,d,} Top view of the isofrequency curves shown in \cref{fig:3}a,c. Arrow direction shows direction of increasing frequency. 
	}
	\label{fig:3}
\end{figure}

In summary, we have demonstrated that plasmonic gain may be achieved by applying an electric current bias to materials with a type-II Dirac node band stucture. While we have used parameters that were fitted to the band structure of WTe$_2$ for most results presented in this work, the gain phenomena observed is generic and can be anticipated for semimetals with an electron hole pocket separation on the order of the plasmon wavevector. General requirements for observing gain are as follows. 
The material must have electron and hole pockets with a momentum space separation on the order of the plasmon wave vector and the current must be applied along the direction of separation. As a result, gain can be imparted to plasmons when its dispersion overlaps with the current driven population inverted single particle transition phase space. We have shown in \cref{fig:1} that for the type-II Dirac node, while both intra- and interband transitions allow for plasmon emission, the intraband transitions are well separated from the plasmon dispersion in the $(\mathbf{q},\omega)$ space. The interband transitions on the other hand overlap with the plasmon dispersion.
Also, recall that contributions from both the $\eta=+1$ and $\eta=-1$ valleys must be included. A given plasmon emission process $(\mathbf{q},\omega)$ in valley $\eta=+1$ may correspond to plasmon absorption in valley $\eta=-1$ which overall cancels out the effect of gain to the plasmon. Hence the emission process that imparts gain to the plasmon must not have an absorbing counterpart in the opposite valley. When these conditions are met, we have shown that internal plasmonic gain may be imparted, thus providing a convenient route towards overcoming the intrinsic loss in the plasmon. The current density required to achieve gain is found to be on the order of a few mA$\mu m^{-1}$ which is achievable in monolayer graphene\cite{Son2018a} and thin film WTe$_2$\cite{Mleczko2016}. Furthermore, we find that the directionality of the plasmon propagation becomes highly dependent on frequency and current magnitude. Thus our proposed setup presents a simple pathway for overcoming plasmonic loss for tightly confined two-dimensional plasmons while also providing control over their propagation direction. Propagation of these plasmons should be readily observable using a typical near-field scanning optical microscope setup\cite{Dong2021a}.

\begin{acknowledgements}
	All authors acknowledge support by the National Science Foundation, NSF/EFRI Grant No. EFRI-1741660.
\end{acknowledgements}

\clearpage
\onecolumngrid

\FloatBarrier
\bibliography{library}
\bibliographystyle{naturemag}
\end{document}